\documentclass[aps,pra,amssymb, amsmath,nobibnotes,nobibnotes,reprint,superscriptaddress,showpacs,showkeys]{revtex4-1}
\usepackage{graphicx}
\usepackage{epstopdf}
\usepackage{graphicx,graphics,color,epsfig}
\usepackage{revsymb4-1}
\usepackage{bm}
\usepackage{braket}
\usepackage[colorlinks, linkcolor=blue, anchorcolor=blue, citecolor=blue]{hyperref}

\begin{document}

\title{Delayed transfer of entanglement to initially populated qubits}

\author{Smail Bougouffa}
\email{sbougouffa@hotmail.com and sbougouffa@imamu.edu.sa}
\affiliation{Department of Physics , College of Science, Imam Mohammad ibn Saud Islamic University (IMSIU), P.O. Box 90950, Riyadh 11623, Saudi Arabia\\
ORCiD:  http://orcid.org/0000-0003-1884-4861}
\author{Zbigniew Ficek}
\email{ficekkacst@gmail.com and z.ficek@if.uz.zgora.pl}
\affiliation{Quantum Optics and Engineering Division, Institute of Physics, University of Zielona G\'ora, Szafrana 4a, Zielona G\'ora 65-516, Poland\\
ORCiD:  http://orcid.org/0000-0002-5260-0169}

\date{\today}

\begin{abstract}
The transfer of entangled, quantum correlated, and flying photons from a squeezed field to single-mode cavities is investigated.
It is shown that, while the transfer of photons begins immediately after the input squeezed field is turned on, the time at which quantum correlations start to be transferred to the cavities is strongly dependent on the initial population of the cavities. For the initially empty cavities, the transfer of quantum correlations begins immediately after the squeezed field is turned on, but it is delayed by a certain time interval when the cavities are initially populated. We find that the transfer of the quantum correlations is postponed until the one-photon states of the system are almost completely depopulated. In other words, the system "waits" for the population of the single-photon states to decay out before starting to build up the quantum correlation between the cavities.
The delay time interval is independent of the number of photons initially present in the system, but is dependent on the decay rates of the cavities and can be varied (controlled) when the cavities decay with different rates. 
It is shown that the delayed transfer of the quantum correlation is directly related to the presence of quantum jumps which transfer the population from the entangled to incoherent mixture states.
\end{abstract}



\maketitle

\section{Introduction}\label{sec1}

Transfer of correlated (entangled) photons from optical beams to stationary quantum systems such as atoms, quantum dots, cavities, and superconducting circuits represents a fundamental problem in quantum information, interferometry, optical communication, and quantum computing~\cite{jk08,hs10,dm10,nb14}. Optical beams of entangled photons such as squeezed light appear as quantum channels, while the stationary systems appear as storage nodes to which quantum states of flying photons are transferred through the mapping process~\cite{km97,hs99,jc00,fg02,sj02,mk04,hh05,yg15}.

Recent theoretical and experimental work on the transfer and storage (mapping) of entanglement or quantum states is focused, to a great extent, on methods and techniques of achieving a significant improvement of the transfer and storage efficiency~\cite{th15,yw16,vh18,w19,ch20,hh17}. Particularly efficient for transfer of a quantum state are systems involving optomechanical cavities with modulated damping rates, in which almost prefect transfer efficiency can be achieved~\cite{wc12,se15,sm15}. 
In the transfer of entanglement, the efficiency relies on the achievement of a large efficiency of mapping the quantum state of the field on states of a stationary system. With the development of quantum processors and interfaces between stationary systems and optical beams, the flying photons can be absorbed with almost perfect efficiency. However, the transfer of a quantum state still be imperfect due to the behavior of the transmitted quantum state as a loss mechanism~\cite{gl87,dk12}. 
This suggests that the manner in which the quantum correlations are transferred is different from that in which the photons are absorbed.
When a system is illuminated by an external field which is composed of uncorrelated photons, eg. a laser field, the photons are absorbed by the system immediately after the field is turned on. A question then arises concerning the transfer of correlated photons. If the external field is composed of correlated photons, are the quantum correlations transferred with the absorbed photons immediately after the field is turned on?

It is the purpose of this paper to address this question by investigating the problem of the transient buildup of quantum correlations (entanglement) between two single-mode cavities exposed to an external source of squeezed light. We are particularly interested in determining how the transfer process of the quantum correlations depends on the initial conditions of the cavities.
We assume that in the time period before $t=0$ the cavities were independent of each other, or equivalently, unentangled. At the time $t=0$, a squeezed vacuum field is applied to the cavities such that each of the output beams drives only one cavity. The transfer of the quantum correlations is then monitored as a function of time and the initial conditions of the cavities. 
We find that the response time of the cavities to transfer the quantum correlations is sensitive to the presence of uncorrelated photons in the system that the time at which the quantum correlations start to be transferred strongly depends on the initial conditions of the cavities. If the initial state of the cavities is the vacuum the transfer of the quantum correlations begins immediately after the squeezed field is turned on. However, the time can be delayed if the cavities are initially in some excited state. 
In other words, the presence of an initial population in the cavities results in a delay in the transfer of quantum correlations. Hence, entanglement can be transferred to the cavities immediately after the squeezed field is turned on if the state of the cavities is the vacuum. The delay time interval can be varied when the cavities decay with different rates. 
We restrict our considerations here to the transfer of the quantum correlations to optical cavities, although similar considerations will apply to other systems such as trapped atoms,
quantum dots or superconducting circuits~\cite{LY00,LS09,MF17,MS20,RG18,W17,te16}. 

The paper is organized as follows. In Sec.~\ref{sec2} we describe the model and introduce the master equation of a pair of independent single-mode cavities illuminated by the output field of a non-degenerate parametric oscillator operating below the threshold. The dynamics of the coherences and populations of the energy states of the cavities for arbitrary initial conditions are discussed in Sec.~\ref{sec3}. We work in the Hilbert space of the system truncated at the energy levels corresponding to two excitations present in the system. Following that discussion, we display the logarithmic negativity, a measure of entanglement, as a function of time for a number of initial conditions.  
The physical interpretation of the results, based on the quantum jumps picture is given in Sec.~\ref{sec4}. In Sec.~\ref{sec5} we conclude with a discussion of our results.

\section{Description of the model}\label{sec2}

\begin{figure}[ht]
\center{\includegraphics[width=0.7\linewidth]{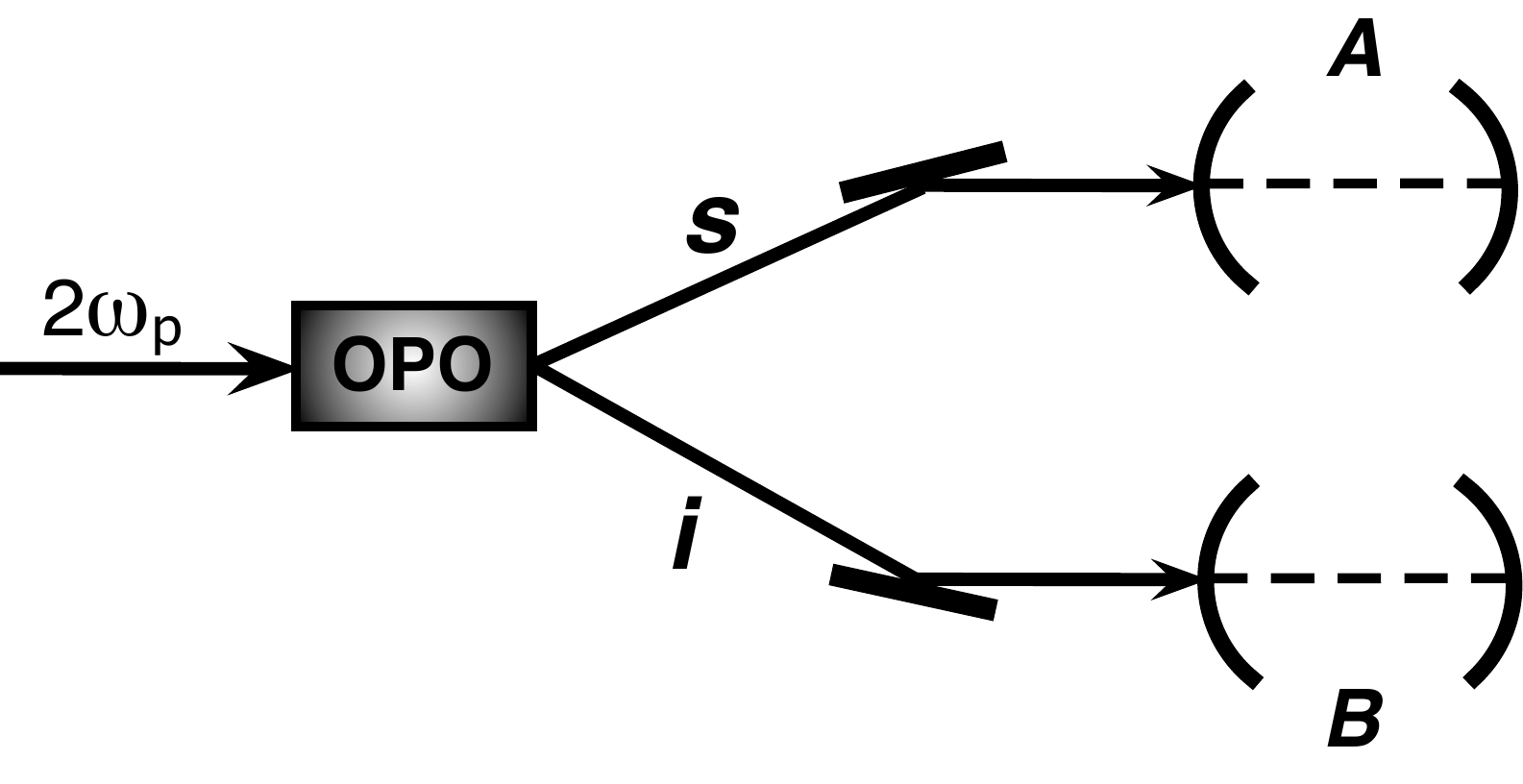}}
\caption{(Color online) Schematic diagram of the system to map entangled state of the output field of an OPO on a pair of single-mode cavities $A$ and $B$. The output signal and idler beams of the OPO driven by a laser of frequency $2\omega_{p}$ are separated spatially and then each of the beams is directed onto one of the cavities. The signal beam illuminates the cavity $A$ and the idler beam illuminates the cavity $B$.}
\label{Fig1}
\end{figure}

We consider a pair of single-mode and single-sided cavities exposed to the output field of an optical degenerate or non-degenerate parametric oscillator (OPO), as shown in Fig.~\ref{Fig1}. In the OPO, the laser beam of frequency $2\omega_{p}$ interacting with a nonlinear medium splits in the process of spontaneous parametric down-conversion into two lower frequency beams, signal $(s)$ and idler $(i)$ beams~\cite{wk86a}. Photons in the idler and signal beams are pretty well localized in space and time, which implies that they are entangled. We assume that the directions of the signal and idler wave vectors do not overlap that the beams were separated spatially, for example, by a prism or polarizer. After the separation, each of the beams is directed onto one of the cavities, such that the signal beam is injected into the cavity $A$, and the idler beam is injected into the cavity $B$. Our problem then is to trace the time evolution of the mapping of flying photons and quantum correlations (entanglement) on the cavities starting from $t=0$ to the steady-state and its dependence on the initial state of the cavities. 

Our starting point is to determine the master equation for the density operator $\rho(t)$ describing the dynamics of the cavities coupled to a broad reservoir field. Under the Born (weak coupling) approximation, the density operator $\rho$ obeys the equation of motion, which in the interaction picture is of the form:
\begin{align}
\frac{d\tilde{\rho}(t)}{dt} &= -\int_{0}^{\infty}dt^{\prime}{\rm Tr}_{R}\left[H_{I}(t),[H_{I}(t-t^{\prime}),\tilde{\rho}(t^{\prime})\otimes\rho_{R}]\right] ,\label{e5}
\end{align}
where $H_{I}(t)$ is the interaction Hamiltonian of the modes of the cavities with the reservoir field, and trace is taken over the reservoir modes. In writing the Hamiltonian (\ref{e5}), we have assumed that the effect of the cavities on the reservoir is very small so that the state of the reservoir remains unchanged. Therefore, the density operator of the system can be assumed to be a tensor product of the time-dependent density operator of the cavities $(\tilde{\rho})$ and the stationary time-independent density operator of the reservoir $(\rho_{R})$. In the rotating-wave approximation, the interaction Hamiltonian $H_{I}$ can be written as $(\hbar=1)$:
\begin{align}
H_{I}(t) &=\!\sum_{j=A,B}\!\int dk\Big[g_{j}(\omega_{k})a_{j}(b_{k}^{\dag}+c_{k}^{\dag})e^{i(\omega_{k}-\omega_{j}\!)t}\!+\!{\rm H.c.}\Big] ,\label{e6}
\end{align}
where $\omega_j$ is the resonance frequency of cavity $j$, $a_{j}$ and $a^{\dag}_{j}$ are the bosonic annihilation and creation operators of cavity~$j$, and $g_{j}(\omega_{k})$ is the coupling constant between the mode of the $j$th cavity and the $k$th mode of the reservoir field. In Eq.~(\ref{e6}), the reservoir has been divided into two parts, one consisting of modes $(b_{k})$ filled with the output field of the OPO and the other modes~$(c_{k})$ being in the ordinary vacuum state. 

The evaluation of the trace over the reservoir modes requires the knowledge of the state of the reservoir modes. The state is determined by correlation functions, which describe the number of photons in the modes and correlations between them. For the modes $c_{k}$ which are in the ordinary vacuum state, the non-zero correlation function~is 
\begin{align}
\langle c_{k}c^{\dagger}_{ k^{\prime}}\rangle &= \delta(\omega_{k} -\omega_{k^{\prime}})  ,\label{e1a}
\end{align}
whereas for the modes $b_{k}$ which are filled with the output field of the OPO the non-zero correlation functions are~\cite{cg85,ga86,df04}
\begin{align}
\langle b_{k}b^{\dagger}_{ k^{\prime}}\rangle &= \left[N(\omega_{k})+1\right]\delta(\omega_{k} -\omega_{k^{\prime}}) ,\nonumber\\
\langle b_{k}^{\dagger}b_{ k^{\prime}}\rangle &= N(\omega_{k})\delta(\omega_{k} -\omega_{k^{\prime}}) ,\nonumber\\
\langle b_{k}b_{k^{\prime}}\rangle &= M(\omega_{k})\delta(2\omega_{p}-\omega_{k}-\omega_{k^{\prime}}) ,\label{e1}
\end{align}
where the frequency-dependent parameter $N(\omega_{k})$ describes the number of photons in the mode of frequency $\omega_{k}$ and $M(\omega_{k})$ describes the degree two-photon correlations between modes of frequencies $\omega_{k}$ and $2\omega_{p}-\omega_{k}$. The parameters can have different forms depending on whether they result from the output field of a degenerate or non-degenerate OPO. In the case of a degenerate OPO, both signal and idler beams are centered at the same frequency $\omega_{p}$, and we have
\begin{align}
N(\omega_{k}) &= \frac{\lambda^{2}-\mu^{2}}{4}\left(\frac{1}{\bar{\omega}_{k}^{2} +\mu^{2}} -\frac{1}{\bar{\omega}_{k}^{2}+\lambda^{2}}\right) ,\nonumber\\
M(\omega_{k}) &= \frac{\lambda^{2}-\mu^{2}}{4}\left(\frac{1}{\bar{\omega}_{k}^{2} +\mu^{2}} +\frac{1}{\bar{\omega}_{k}^{2}+\lambda^{2}}\right) ,\label{e2}
\end{align}
where $\bar{\omega}_{k}=\omega_{k} -\omega_{p}$,  $\mu =\frac{1}{2}\kappa_{c}-\varepsilon$ and $\lambda =\frac{1}{2}\kappa_{c}+\varepsilon$. In the parameters $\mu$ and $\lambda$, $\kappa_{c}$ is the damping constant of the OPO cavity, and $\varepsilon$ is its amplification parameter, proportional to the amplitude of the pumping field.

In the case of a non-degenerate OPO, signal and idler beams are centered at different frequencies $(\omega_{p}\pm \alpha)$ displayed from $\omega_{p}$ by $\alpha$, and we have 
\begin{align}
N(\omega_{k}) &= \frac{\lambda^{2}-\mu^{2}}{4}\!\left[\frac{1}{(\bar{\omega}_{k} -\alpha)^{2}\!+\!\mu^{2}} -\frac{1}{(\bar{\omega}_{k} -\alpha)^{2}\!+\!\lambda^{2}}\right. \nonumber\\
&\left. +\frac{1}{(\bar{\omega}_{k}+\alpha)^{2}\!+\!\mu^{2}} -\frac{1}{(\bar{\omega}_{k}+\alpha)^{2}\!+\!\lambda^{2}}\right] ,\nonumber\\
M(\omega_{k}) &= \frac{\lambda^{2}-\mu^{2}}{4}\!\left[\frac{1}{(\bar{\omega}_{k} -\alpha)^{2}+\mu^{2}} +\frac{1}{(\bar{\omega}_{k} -\alpha)^{2}\!+\!\lambda^{2}}\right. \nonumber\\
&\left. +\frac{1}{(\bar{\omega}_{k}+\alpha)^{2}\!+\!\mu^{2}} +\frac{1}{(\bar{\omega}_{k}+\alpha)^{2}\!+\!\lambda^{2}}\right] . \label{e3}
\end{align}

It is easily shown that the degree of two-photon correlations $M(\omega_{k})$ depends on the number of photons in the modes, such that
\begin{equation}
M^{2}(\omega_{k}) = N(\omega_{k})\left[N(2\omega_{p}-\omega_{k})+1\right] .\label{e4}
\end{equation}

The parameters (\ref{e2}) and (\ref{e3}) are Lorentzian type functions with amplitudes and widths determined by $\mu$ and $\lambda$, which can be varied by varying the OPO parameters $\kappa_{c}$ and $\varepsilon$. The limit $\lambda,\mu\rightarrow\infty$ such that $\lambda/\mu ={\rm const.}$, corresponds to an infinitely broad (frequency-independent) squeezed vacuum field. For finite values of $\mu$ and $\lambda$, but $\lambda,\mu \gg \kappa_{1},\kappa_{2}$, where $\kappa_{1}$ and $\kappa_{2}$ are the damping rates of the cavities, the OPO output field can be treated as a broadband squeezed vacuum reservoir to the cavity modes~\cite{am13,sy16}. 

We now use the Hamiltonian (\ref{e6}) and the results for the correlation functions, Eqs.~(\ref{e1}) and (\ref{e2}), to calculate the trace of the double commutator appearing in Eq.~(\ref{e5}). Assuming that $\lambda$ and $\mu$ are much larger than the damping rates of the cavities, we can make the Markov approximation in which we replace $\tilde{\rho}(t^{\prime})$ by $\tilde{\rho}(t)$, so that we can extract $\tilde{\rho}(t)$ from the integral. The integral then can be evaluated and we arrive to the following master equation~\cite{KSZ14, BH11, BF16}
\begin{eqnarray}
\frac{d\tilde{\rho}(t)}{dt} &=& -\sum_{i,j}\frac{1}{2}\kappa_{ij}\Bigg\{\eta N(\omega_{j})\left([a_{i},a_{j}^{\dag}\tilde{\rho}] + [\tilde{\rho} a_{i},a_{j}^{\dag}]\right) \nonumber\\
&&+[\eta N(\omega_{j})+1]\left([a_{j}^{\dag},a_{i}\tilde{\rho}] + [\tilde{\rho} a_{j}^{\dag},a_{i}]\right)\!\Bigg\}e^{i\Delta_{ij}t} \nonumber\\
&-&\sum_{i\neq j}\frac{1}{2}\kappa_{ij}\Bigg\{ \eta M\!(\omega_{j})\Big(\!\left[a_{i}\tilde{\rho},a_{j}\right]\!+\!\left[a_{j},\tilde{\rho} a_{i}\right]\!\Big)e^{i\Delta_{p}t} \nonumber\\
&&+\eta M^{\ast}(\omega_{j})\Big([a_{i}^{\dag}\tilde{\rho},a_{j}^{\dag}] +[a_{j}^{\dag},\tilde{\rho} a_{i}^{\dag}]\Big)e^{-i\Delta_{p}t}\Bigg\} ,\label{e7}
\end{eqnarray}
where $\Delta_{ij}=\omega_{i}-\omega_{j}$, $\Delta_{p} =2\omega_{p}-\omega_{i}-\omega_{j}$, and $\eta$ is the efficiency with which the OPO field couples to the cavities. 
The parameters $\kappa_{ij}$ are related to damping in the cavity system which, on the other hand, depends on the coupling constants of the cavity modes to the reservoir modes~\cite{lo73,ag74,wk86} 
\begin{equation}
\kappa_{ij} =\pi g_{i}(\omega_{i})g_{j}^{\ast}(\omega_{j}) .
\end{equation}
Here, $\kappa_{ii}\equiv \kappa_{i}$ describes the damping rate of cavity $i$, and $\kappa_{ij}\ (i\neq j)$ the damping of cavity $i$ caused by the output field of cavity $j$.  
It is easily seen that the terms describing the incoherent damping $(\eta N+1)$ and incoherent pumping $(\eta N)$ processes involve both the $i=j$ and $i\neq j$ terms. However, the two-photon correlation terms, proportional to $M$, involve only the $i\neq j$ terms. It is a reflection of the fact that the idler and signal beams are in a thermal state with no correlations between photons inside each beam. The correlations are present only between the beams, as indicated by the correlation functions of the reservoir modes, Eq.~(\ref{e1}).

The master equation (\ref{e7}) describes the dynamics of the collectively decaying cavities, which results from the presence of the cross damping rate $(\kappa_{12})$. However, the collective decay of the cavities may create entanglement even if the input field is in the ordinary vacuum state. Thus, to see effects, which are related solely to correlations present in the input field rather than caused by the correlations inside the system, it is better to work in the regime the collective effects are absent that the cavities decay independent of each other. In this case, any correlations between cavities will correspond to those transferred from the input field.

To achieve this situation, it is sufficient to work with cavities of significantly different frequencies that $\Delta_{ij}$ exceeds the bandwidth of the cavity modes but remains inside the bandwidth of the input field. In this case, we can neglect the cross-terms in the dissipative part of the master equation. This simplification retains the two-photon correlation terms, which in the case of $\omega_{p}=\omega_{0}\ (\Delta_{p}=0)$ become independent of time. 
Hence, in the absence of the input two-photon correlations $(M=0)$, the cavities decay independently without any dynamical influence on one another through the background (reservoir) field. 

Under the assumption that $\Delta_{ij}\gg \kappa_{ij}$, the master equation (\ref{e7}) reduces to 
\begin{eqnarray}
\frac{d\tilde{\rho}(t)}{dt} &=& -\sum_{j=A,B}\frac{1}{2}\kappa_{j}\Bigg\{\eta N(\omega_{j})\left([a_{j},a_{j}^{\dag}\tilde{\rho}] +[\tilde{\rho} a_{j},a_{j}^{\dag}]\right) \nonumber\\
&&+[\eta N(\omega_{j})+1]\left([a_{j}^{\dag},a_{j}\tilde{\rho}] +[\tilde{\rho} a_{j}^{\dag},a_{j}]\right)\Bigg\} \nonumber\\
&-&\sum_{i\neq j=A,B}\frac{1}{2}\kappa_{ij}\Bigg\{ \eta M(\omega_{j})\Big(\left[a_{i}\tilde{\rho},a_{j}\right]\!+\!\left[a_{j},\tilde{\rho} a_{i}\right]\Big) \nonumber\\
&&+\eta M^{\ast}(\omega_{j})\Big([a_{i}^{\dag}\tilde{\rho},a_{j}^{\dag}] +[a_{j}^{\dag},\tilde{\rho} a_{i}^{\dag}]\Big)\Bigg\} .\label{e8}
\end{eqnarray}

Before moving on to the consideration of the response time of the cavities to the quantum correlations present in the input squeezed field, we first estimate the values of the parameters which are experimentally convenient and at which the output modes of the OPO are strongly entangled. It is well known, that the output modes are entangled when $M-N>0$, and the maximal value of $M$ is $\sqrt{N(N+1)}$ which corresponds to the maximum degree of squeezing (entanglement) possible for a given $N$. It is easily verified that the relative difference $M-N$ is very small for $N\gg 1$, but is very large for $N\ll 1$. We therefore will largely concentrate on small values of $N$ and choose $N=0.125\ (M=0.375)$. This choice of the parameters corresponds to the degree of squeezing used in the experiments on spectroscopy with squeezed light~\cite{te16,tg98}. These experimental values correspond to $\lambda =\sqrt{2}\mu$, i.e., a weak pumping with $\epsilon =0.17\kappa_{c}/2$, giving
\begin{equation}
\lambda = \frac{\sqrt{2}}{\sqrt{2}+1}\kappa_{c} ,\quad \mu = \frac{1}{\sqrt{2}+1}\kappa_{c} ,
\end{equation}
at which the output OPO field exhibits $50\%$ squeezing~\cite{cg85,pa93,ag16}. Moreover, it shows that to satisfy the requirement that the output OPO field is broadband compared to the damping rates of the cavities, the decay rate of the OPO cavity $\kappa_{c}$ should be larger than the decay rates of the cavities $A$ and $B$, $\kappa_{c}\gg \kappa_{1},\kappa_{2}$. In the OPO systems bandwidths up to $5.8$ MHz have been achieved~\cite{tg98}, whereas in a Josephson parametric amplifier (JPA) bandwidths up to $38$ MHz~\cite{te16} have been achieved. 
It should be pointed out that this is the only condition imposed on the damping rates involved in master Eq. (\ref{e8}), i.e., no further adjustment of the damping rates to experimental values is required.

Since we consider cases of a weak excitation $(N\ll 1)$, we will restrict our calculations to a limited basis set by truncating the Hilbert space of the system at two-photon states. 
The system then has six energy states
\begin{align} 
\ket{1} &= \ket{0_{A}0_{B}}, \ket{2}= \ket{1_{A}0_{B}}, \ket{3}=\ket{0_{A}1_{B}} ,\nonumber \\
\ket{4} &= \ket{1_{A}1_{B}}, \ket{5}= \ket{2_{A}0_{B}}, \ket{6}=\ket{0_{A}2_{B}} .\label{b1}
\end{align}
The zero photon state $\ket 1$ is a singlet, the single-photon states $\ket 2$ and $\ket 3$ form a degenerate doublet, and the two-photon states $\ket 4, \ket 5$ and $\ket 6$ form a degenerate triplet, as shown in Fig.~\ref{Fig6}. 

The annihilation operators of the cavity modes can now be expressed in terms of the projection operators between the basis states
\begin{align}
\hat{a}_{A} &= \ket 1\bra 2 +\sqrt{2}\ket 2\bra 5 +\ket 3\bra 4 ,\nonumber\\
\hat{a}_{B} &= \ket 1\bra 3 +\sqrt{2}\ket 3\bra 6 +\ket 2\bra 4 .\label{b2}
\end{align}
With the help of the master equation (\ref{e7}) equations of motion can be set up for the populations of the cavities and coherences between them, 
and solved to study the evolution in time of the cavity system.

\section{Transfer of quantum correlations to the decaying cavities}\label{sec3}

We now consider the process of transferring the quantum correlations (entanglement) to the cavities from the input squeezed field.
We assume that in the time before $t=0$ the cavities were independent of each other, or equivalently, unentangled. At the time $t=0$, a squeezed field is applied to the cavities and the transfer of the quantum correlations is then monitored as a function of time and the initial conditions of the cavities.

Let us first check if the cavities decay to the steady-state, which is a pure entangled state, and that the correlations induced between the cavities are only those existing in the input squeezed field. Consider the equations of motion for the density matrix elements, Eq.~(\ref{b5}). 
If we introduce incoherent mixtures of the one-photon states $\ket 2$ and $\ket 3$, and also the two-photon states $\ket 5$ and $\ket 6$,
\begin{align}
\rho_{ss} &= \frac{1}{2}\left(\rho_{22} +\rho_{33}\right) ,\quad \rho_{uu} =\frac{1}{2}\left(\rho_{55} +\rho_{66}\right) ,\label{b6}
\end{align}
we then find that the set of coupled equations~(\ref{b5}) reduces to
\begin{align}
\dot{\rho}_{11} =& -2N\kappa\rho_{11} +2(N+1)\kappa\rho_{ss} + 2M\kappa\rho_{m} ,\nonumber\\
\dot{\rho}_{ss} =& -\left(4N+1\right)\kappa\rho_{ss} +(N+1)\kappa\left(\rho_{44} +2\rho_{uu}\right) \nonumber\\
&+ N\kappa\rho_{11} -2M\kappa\rho_{m} ,\nonumber\\
\dot{\rho}_{44} =& -2(N+1)\kappa\rho_{44} +2N\kappa\rho_{ss} + 2M\kappa\rho_{m} ,\nonumber\\
\dot{\rho}_{uu} =& -2(N+1)\kappa\rho_{uu} +2N\kappa\rho_{ss} ,\nonumber\\
\dot{\rho}_{m} =& -(2N+1)\kappa\rho_{m} + M\kappa(\rho_{11}+\rho_{44}-2\rho_{ss}) ,\label{b7}
\end{align}
where $\rho_{m} = \left(\rho_{14} +\rho_{41}\right)/2$.

We see that the evolution of the populations is affected solely by the two-photon coherence $\rho_{m}$. The coherence creates superposition states involving the ground state $\ket 1$ and the doubly excited state $\ket 4$. In the case of a quantum squeezed field with the correlations $M=\sqrt{N(N+1)}$, we may introduce two orthogonal superposition states
\begin{align}
\ket \alpha & =\sqrt{\frac{N+1}{2N+1}}\ket 1 +\sqrt{\frac{N}{2N+1}}\ket 4 ,\nonumber\\
\ket\beta &= \sqrt{\frac{N}{2N+1}}\ket 1 -\sqrt{\frac{N+1}{2N+1}}\ket 4 ,\label{b8}
\end{align}
and find that Eq.~(\ref{b7}) lead to the following equations of motion
\begin{align}
\dot{\rho}_{\alpha\alpha} =& \frac{2\kappa}{2N+1}\rho_{ss} ,\nonumber\\
\dot{\rho}_{\beta\beta} =& -2(2N+1)\kappa\rho_{\beta\beta} +\frac{8N(N+1)}{2N+1}\kappa\rho_{ss} ,\nonumber\\
\dot{\rho}_{ss} =& -\left(4N+1\right)\kappa\rho_{ss} +(2N+1)\kappa\rho_{\beta\beta} +2(N+1)\kappa\rho_{uu} ,\nonumber\\
\dot{\rho}_{uu} =& -2(N+1)\kappa\rho_{uu} +2N\kappa\rho_{ss} .\label{b9}
\end{align}
It is easy to see that in the steady-state, $\rho_{ss}= 0$. As a consequence, $\rho_{uu}=\rho_{\beta\beta} =0$ and then $\rho_{\alpha\alpha}=1$. Thus, in the quantum squeezed field the system
decays to the pure state $\ket\alpha$. It is also obvious that the state $\ket\alpha$ is an entangled state with the degree of entanglement
\begin{equation}
\frac{2\sqrt{N(N+1)}}{2N+1} .
\end{equation}

It is well known that the signal and idler beams of the OPO output field always behave as mutually incoherent that the first-order coherence measured by the cross-correlation $\langle a^{\dagger}_{i}a_{s}\rangle$ of the idler and signal fields $a_{i}$ and $a_{s}$ is zero~\cite{mg96,ma98,wm00,hm15,hr14,hm15a,mh19}. Moreover, the anomalous correlations measured by the correlations $\langle a^{2}_{i}\rangle$ and $\langle a^{2}_{s}\rangle$ are also zero, that the idler and signal beams are each in a thermal state. However, the mutual anomalous correlation measured by the cross-correlation $\langle a_{i}a_{s}\rangle$ is non-zero.  
Before proceeding further, we will check if the cavities also exhibit the same correlation properties.
To check this, we introduce the normalized correlation functions (degrees of coherence), the first-order coherence function
\begin{align}
|\gamma_{12}| =& \frac{|\langle\hat{a}^{\dagger}_{1}\hat{a}_{2}\rangle |}{\sqrt{\langle\hat{a}^{\dagger}_{1}\hat{a}_{1}\rangle\langle\hat{a}^{\dagger}_{2}\hat{a}_{2}\rangle}} ,
\end{align}
and the anomalous coherence functions
\begin{align}
|\eta_{ii}|=& \frac{|\langle a_{i}a_{i}\rangle |}{\langle\ a^{\dagger}_{i}a_{i}\rangle} ,\quad i=1,2 \\
|\eta_{ij}|=& \frac{|\langle a_{i}a_{j}\rangle|}{\sqrt{\langle a^{\dagger}_{i}a_{i}\rangle\langle a^{\dagger}_{j}a_{j}\rangle}} ,\quad i\neq j=1,2 .
\end{align}
In terms of the density matrix elements the coherence functions are given by
\begin{align}
|\gamma_{12}|=&\frac{|\rho_{23}+\sqrt{2}\rho_{54}+\sqrt{2}\rho_{46}|}{\sqrt{(\rho_{22}+2\rho_{55}+\rho_{44})(\rho_{33}+2\rho_{66}+\rho_{44})}} , \\
|\eta_{11}|=&  \frac{\sqrt{2}|\rho_{51}|}{\rho_{22}+2\rho_{55}+\rho_{44}}  , \\
|\eta_{22}|=&  \frac{\sqrt{2}|\rho_{61}|}{\rho_{33}+2\rho_{66}+\rho_{44}} , \\
|\eta_{12}|=& \frac{|\rho_{41}|}{\sqrt{(\rho_{22}+2\rho_{55}+\rho_{44})(\rho_{33}+2\rho_{66}+\rho_{44})}} .
\end{align}

\begin{figure}[h]
\center{\includegraphics[width=0.7\linewidth]{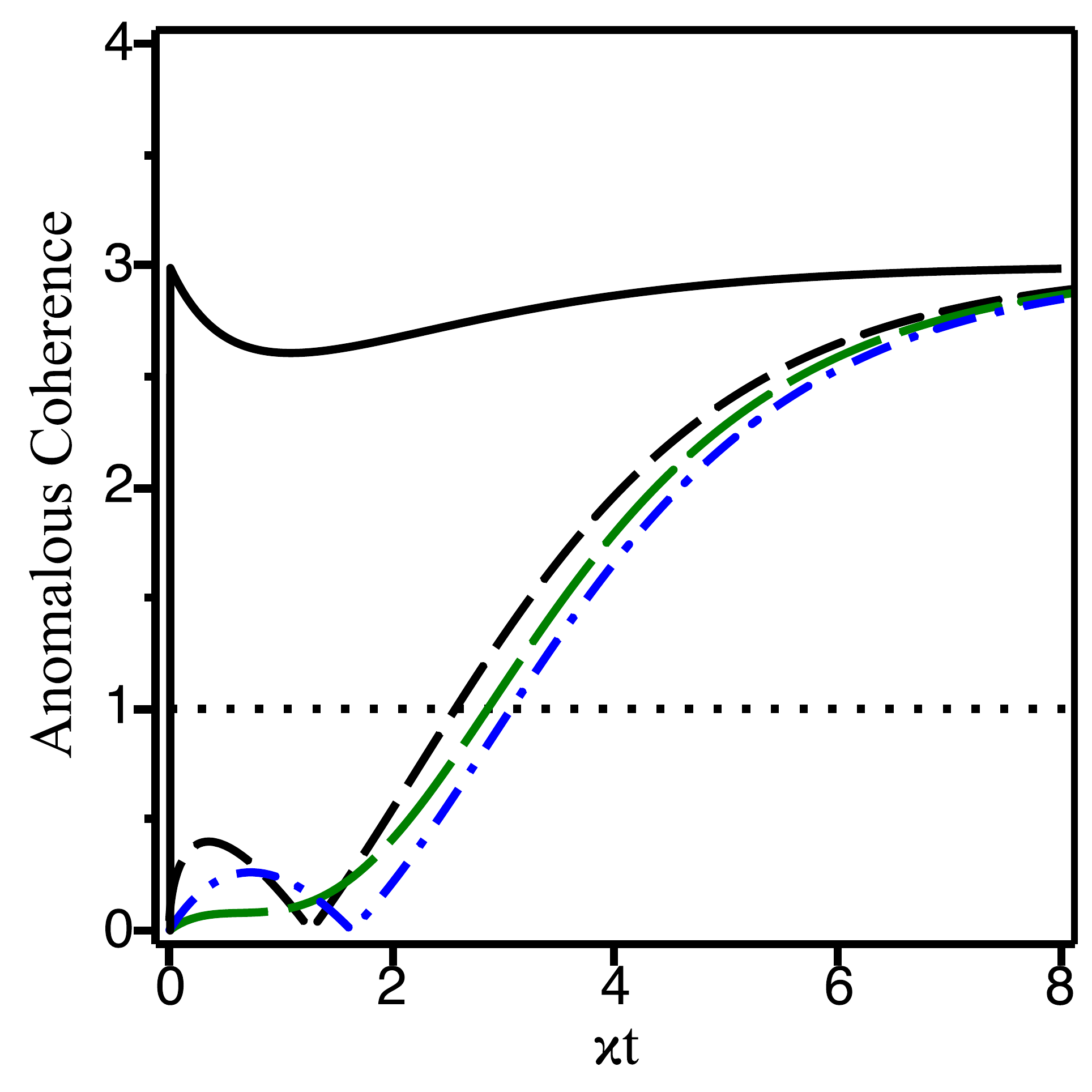}}
 \caption{(Color online) The transient behavior of the anomalous coherence $|\eta_{12}(t)|$ for the initial conditions 
 $\rho_{11}(0)=1$ (black solid line), $\rho_{22}(0)=1$ (black dashed line), $\rho_{44}(0)=1$ (green long dashed line), and $\rho_{55}(0)=1$ (blue dashed-dotted line), with $N=0.125$, $|M|=\sqrt{N(N+1)}$, $\eta=1$ and $\kappa_1=\kappa_2=\kappa_{12}=\kappa$. }
\label{Fig2}
\end{figure}

Assume that in the time before $t=0$ the cavities were prepared in a separable (unentangled) state and then exposed at $t=0$ to the squeezed field. Then, according to the equations of motion (\ref{b9}), the coherences $\rho_{23}, \rho_{54}, \rho_{46}, \rho_{51}$ and $\rho_{61}$ are not coupled to the populations, they are zero in the system of two cavities initially prepared in an unentangled state. Therefore, the first-order coherence function $|\gamma_{12}|$ and the anomalous coherence functions $|\eta_{11}|$ and $|\eta_{22}|$ are zero. The two-photon coherence $\rho_{14}$ is coupled to the populations. Therefore it can be different from zero. This is illustrated in Fig.~\ref{Fig2}, which shows the anomalous coherence $|\eta_{12}|$ as a function of time for different initial conditions. It is evident that the coherence builds up during the evolution of the system and reaches a nonzero steady-state value. Thus, we may conclude that the correlations present between cavities are those transferred from the input squeezed field and that the cavities decay to the steady-state, which is the pure entangled state~\cite{ep89,ap90,ft17}.

One can notice from Fig.~\ref{Fig2} that the transient buildup of the anomalous coherence depends on the initial population of the cavity modes. In other words, the response of the cavities to the input squeezed field at $t=0$ depends on whether the cavities were initially populated or not. 
When the cavities are exposed to the squeezed field, one could expect that the correlation should build up in the cavities immediately after the field is turned on at $t=0$. 
However, it happens only when the cavities are empty, i.e., they are in their ground states. When a population is initially present in the cavities, the buildup of the coherence is delayed by a certain time interval. The time interval is practically independent of which of the cavity excited states is initially populated.
\begin{figure}[h]
\center{{\bf (a)}\hskip3.5cm{\bf (b)}}
\resizebox{0.99\linewidth}{!}{%
\includegraphics{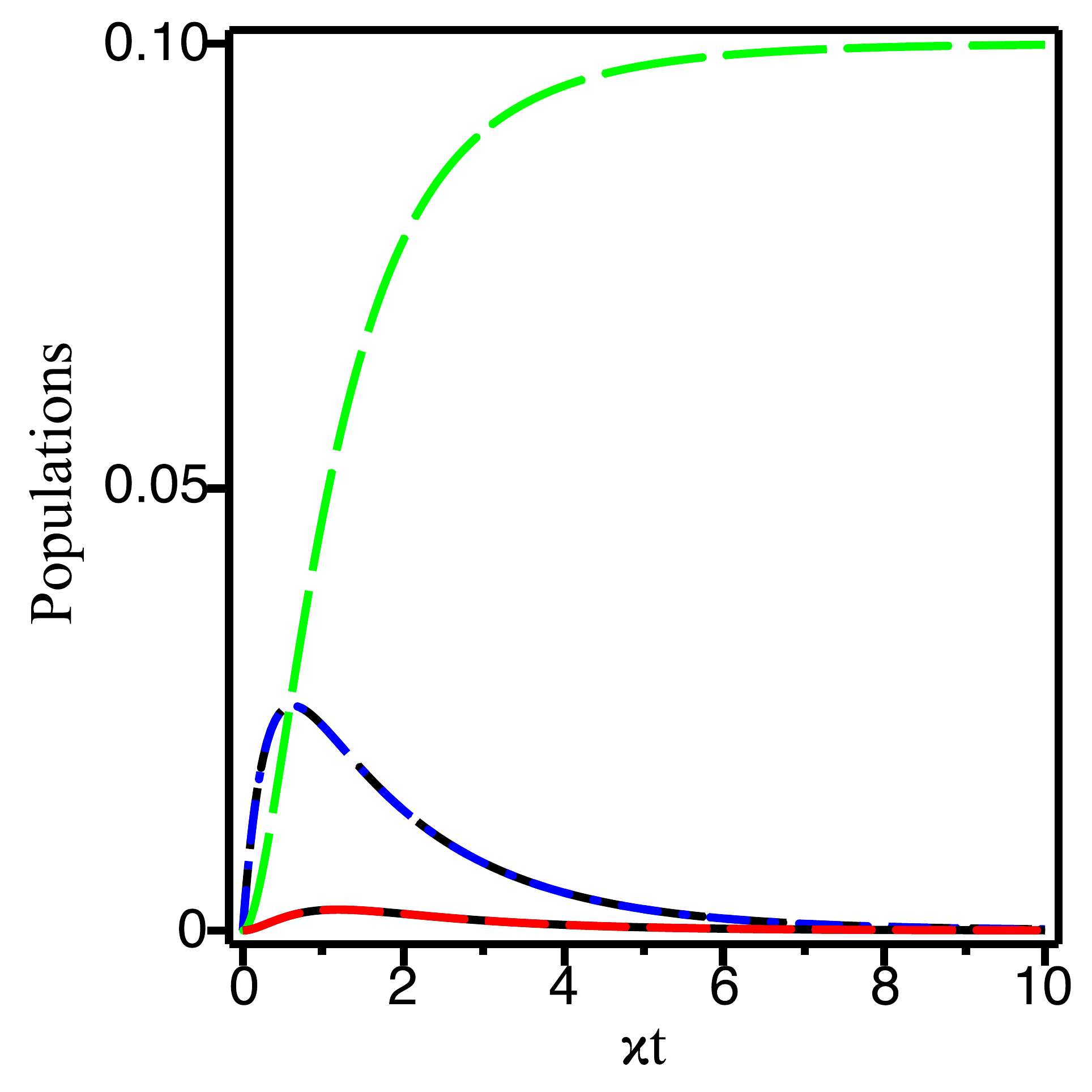}~\includegraphics{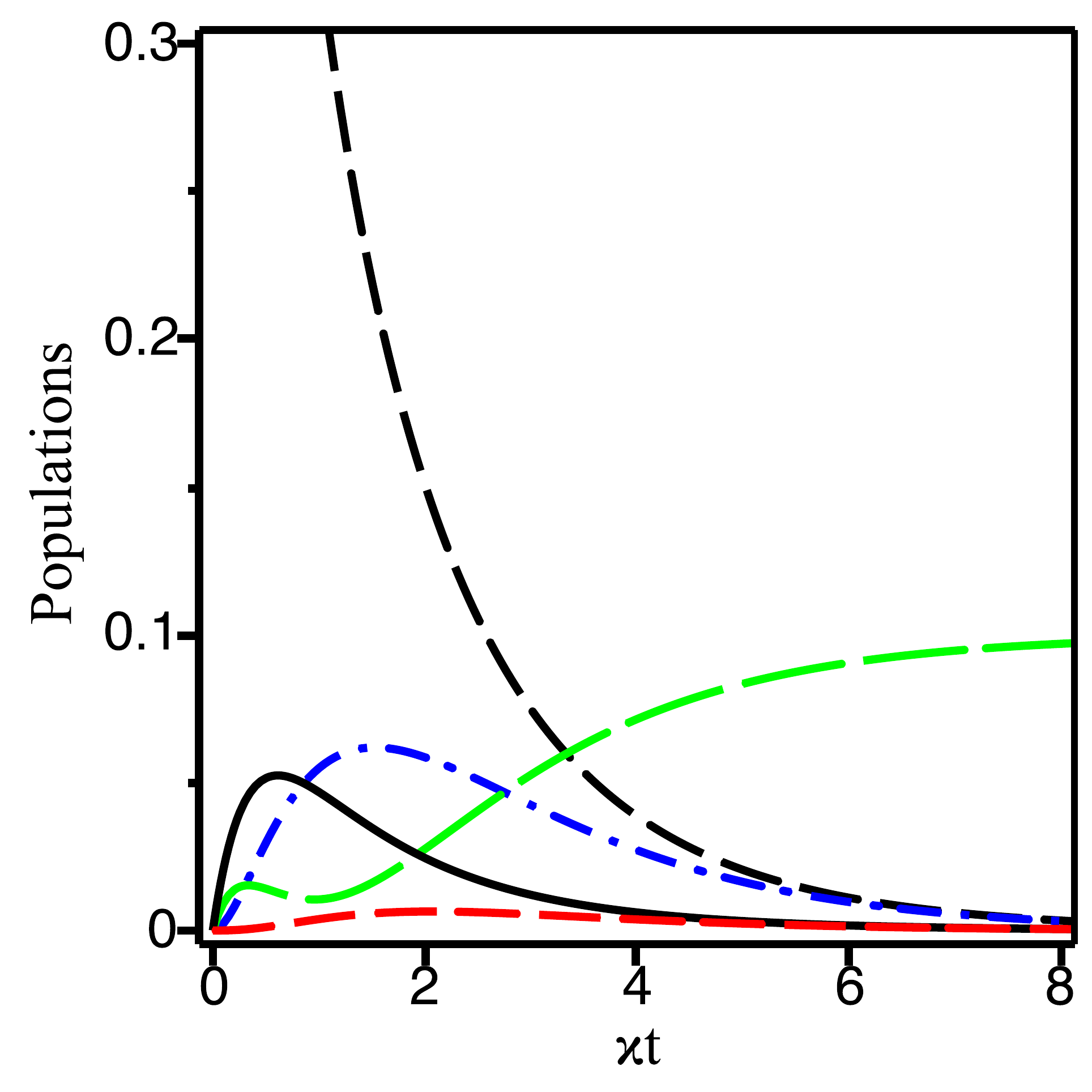}}\\
\center{{\bf (c)}\hskip3cm{\bf (d)}}
\resizebox{0.99\linewidth}{!}{%
\includegraphics{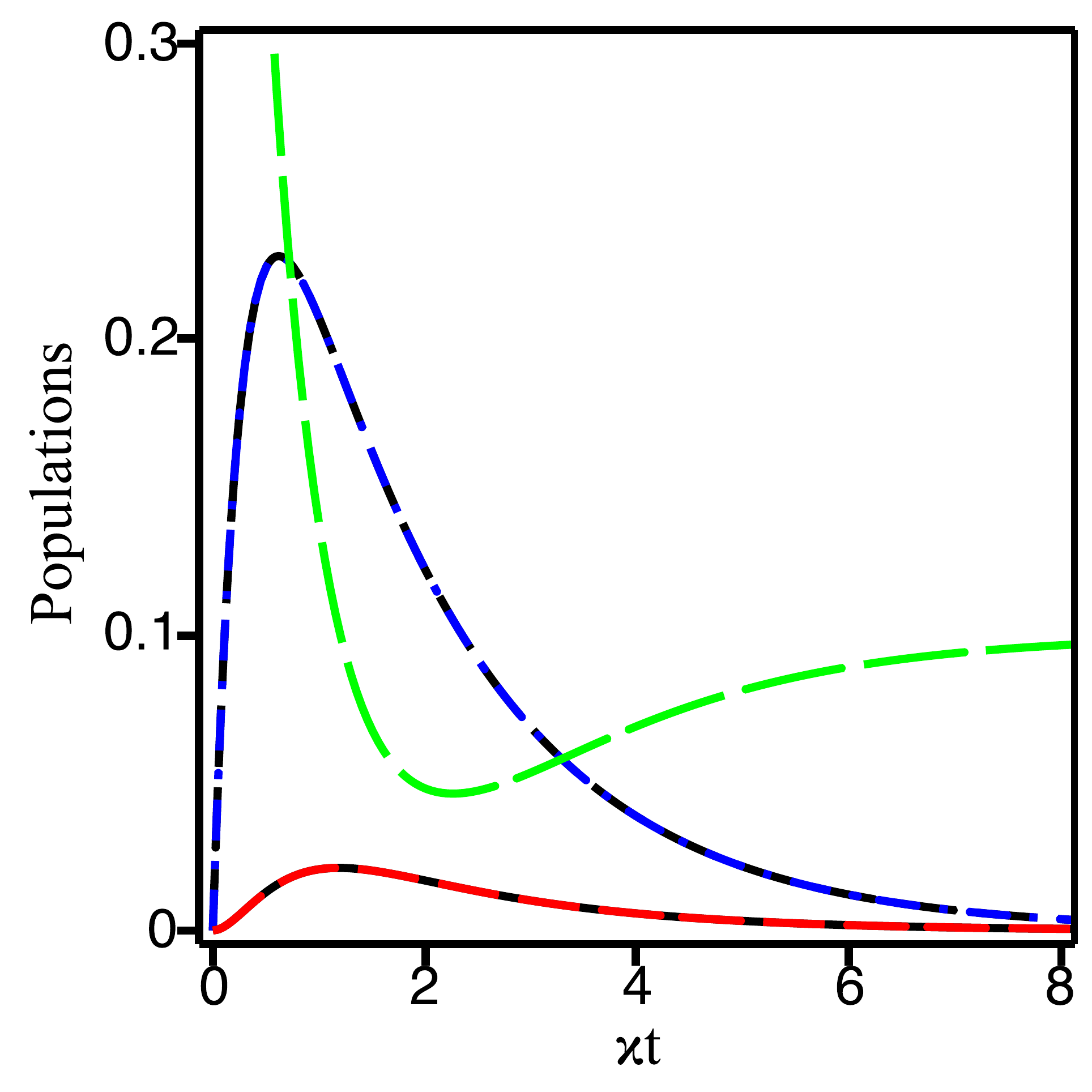}~\includegraphics{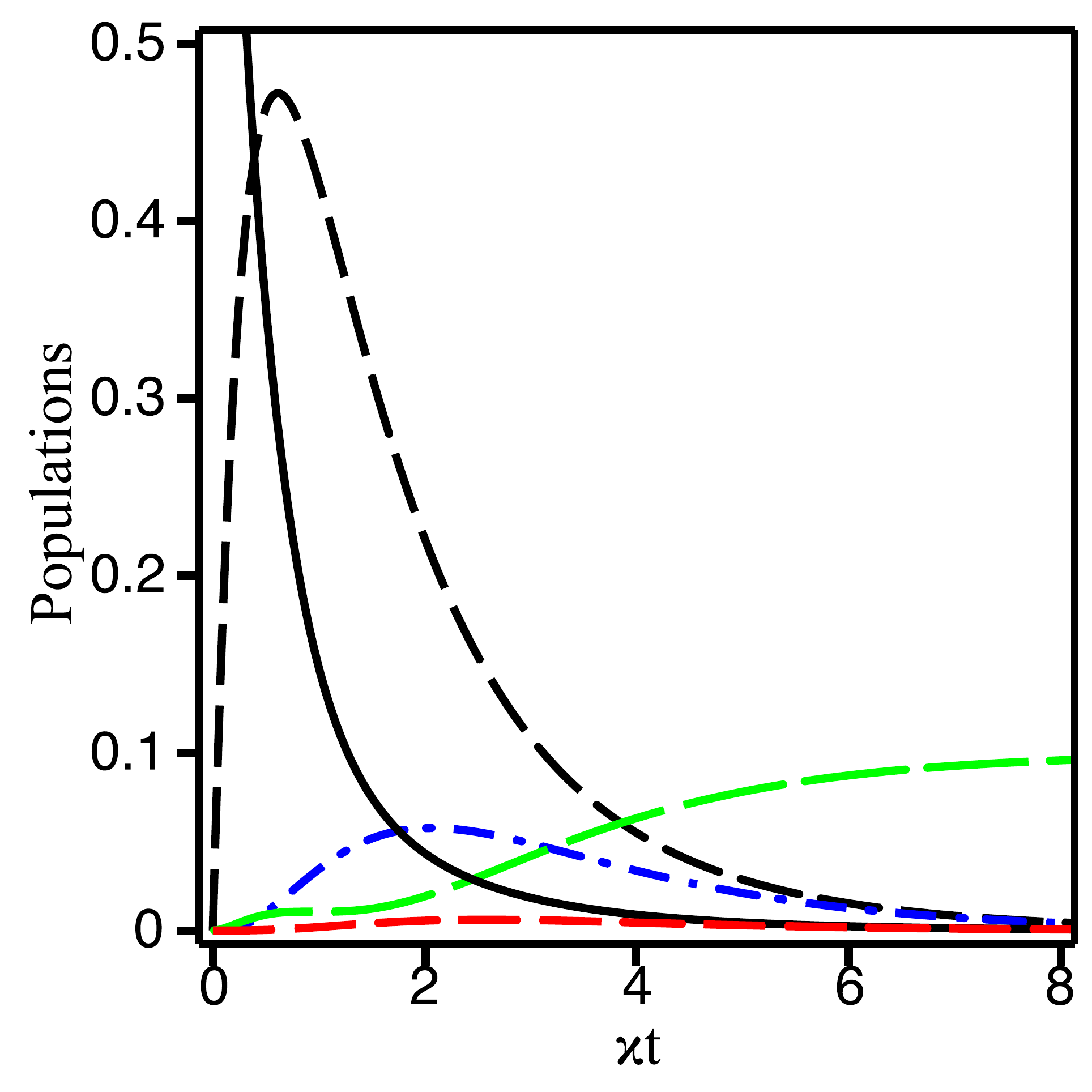}}
 \caption{(Color online) The transient behavior of the populations $\rho_{22}$ (black dashed line), $\rho_{33}$ (blue dashed-dotted line), $\rho_{44}$ (green long dashed line), $\rho_{55}$ (black solid line), and $\rho_{66}$ (red dashed line) when the cavities are initially prepared in the state (a) $\ket{1}$, (b) $\ket{2}$, (c) $\ket{4}$ and (d) $\ket{5}$, with $N=0.125$, $|M|=\sqrt{N(N+1)}$, $\eta=1$ and $\kappa_1=\kappa_2=\kappa_{12}=\kappa$. }
\label{Fig3}
\end{figure}

For comparison, we plot in Fig~\ref{Fig3} the time evolution of the populations of the excited energy states of the cavity system for the same parameters and the initial conditions as in Fig.~\ref{Fig2}. It is seen that independent of the initial conditions, the populations of the initially unpopulated states start to buildup immediately after the squeezed field is turned on at $t=0$. 

It is interesting to note that at early times the population of the single-photon states builds up more rapidly than the population of the two-photon states. Further, we observe that the population of the two-photon state $\ket 4$, which is coupled to the ground state through the two-photon coherence $\rho_{14}$, is suppressed at times the one-photon states are significantly populated. Comparing with Fig.~\ref{Fig2}, we see that not only the population of the state $\ket 4$ but also the anomalous coherence is significantly reduced at times when the single-photon states are populated. Thus, the buildup of the anomalous coherence is strongly affected by the presence of the population in the single-photon states.

The behaviour of the anomalous coherence should be reflected in the behaviour of entanglement between the cavities. To quantify entanglement we adopt the logarithmic negativity,
which for a bipartite system is defined as~\cite{VW02, LCK03}
\begin{equation}
    \mathcal{N}=\log_{2}(1+2|\sum_{l}\mu_{l}|) ,\label{e11}
\end{equation}
where $\mu_{l}$ are the negative eigenvalues of $\rho^{T_{B}}$, the partial transpose of a state $\rho$ in
$n\otimes n'$ ($n\leq n'$) quantum system. For an unentangled state, $\mathcal{N}=0$, whereas $\mathcal{N}=1$ for the maximally entangled state.
\begin{figure}[h]
\center{\includegraphics[width=0.7\linewidth]{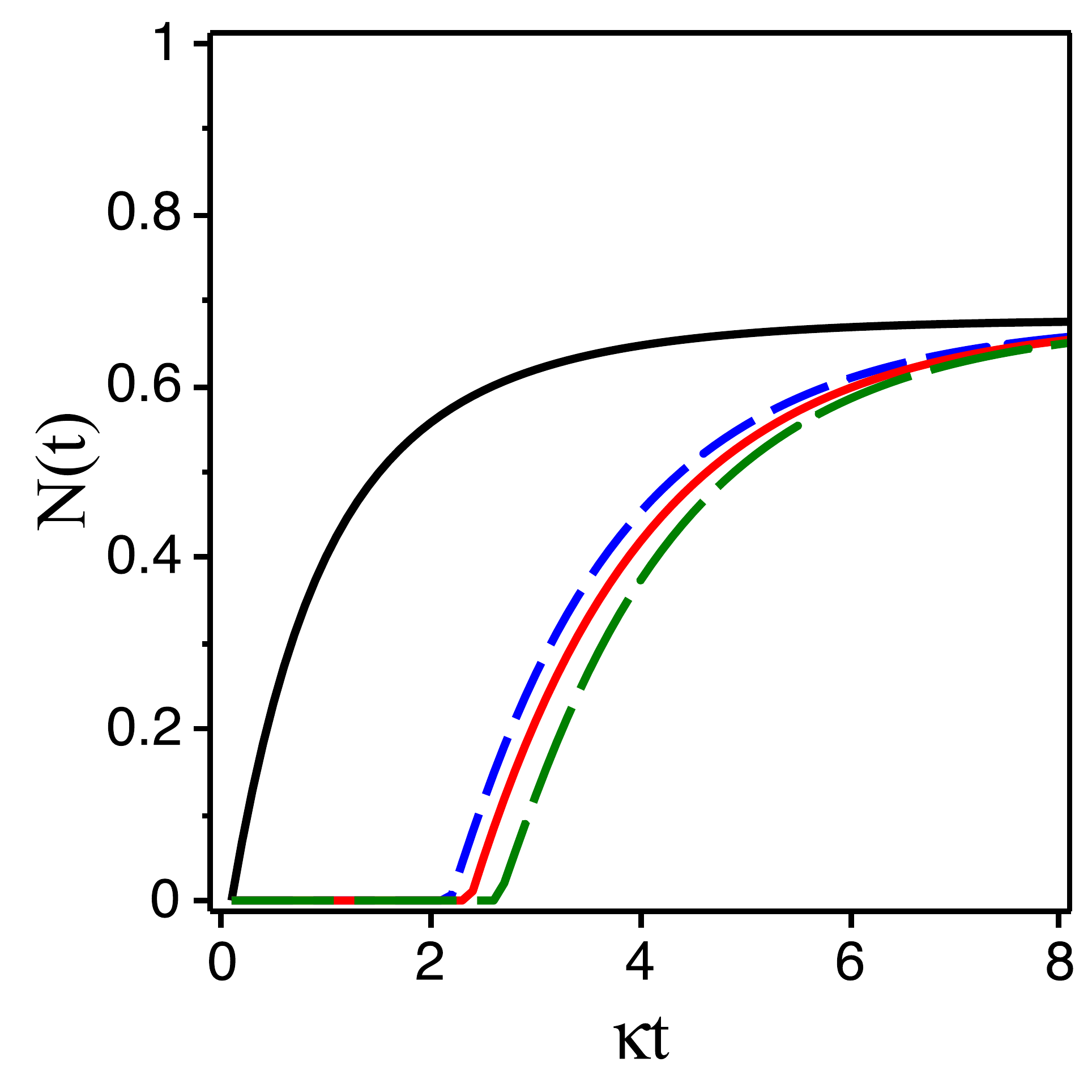}}
\caption{(Color online) Logarithmic negativity as a function of the dimensionless time $\kappa t$ illustrating the transient buildup of entanglement between independently decaying cavities for $N=0.125$, $|M|=\sqrt{N(N+1)}$, $\eta=1$, $\kappa_1=\kappa_2=\kappa_{12}=\kappa$ and different initial conditions: $\rho_{11}(0)=1$ (black solid line), 
$\rho_{22}(0)=1$ (blue dashed line), $\rho_{44}(0)=1$ (red solid line),  and $\rho_{55}(0)=1$ (green long dashed line).}
\label{Fig4}
\end{figure}

Figure~\ref{Fig4} shows the time evolution of the logarithmic negativity of the cavity system for the same parameters and the initial conditions as in Fig.~\ref{Fig2}. 
It is evident that the transient buildup of entanglements exhibits the same behaviour as the anomalous coherence function. For the initial vacuum state, $\rho_{11}(0)=1$, the buildup of the entanglement, ie., transfer of the quantum correlations to the cavities, starts immediately after the input OPO field is turned on. 
The time behaviour of the entanglement appears qualitatively different when cavities are initially prepared in one of their excited states. In this case, the transfer of the quantum correlations is delayed by a finite time interval. We may conclude that the presence of an unentangled photons in the cavities prevents (blocks) the transfer of the quantum correlations from the incident squeezed field. Thus, at early times the transfer of the entanglement from the external field is suppressed. The transfer of the entanglement is delayed to the time $t\approx 2\kappa^{-1}$, as seen from Fig.~\ref{Fig4}. This means that the delay is in the order of the decoherence time of the cavities. 

When comparing the time evolution of the logarithmic negativity, Fig~\ref{Fig4}, with the time evolution of the populations of the cavity states, Fig~\ref{Fig3}, one finds that the transfer of the entanglement and also the build up of the population of the state $\ket 4$ are postponed till the one photon states are almost completely depopulated. 
In other words, the system "waits" for the population of the single photon states to decay out before starting the transfer of the entanglement from the incident squeezed field. 
\begin{figure}[h]
\hskip0.15cm {\bf (a)}\hskip4.0cm{\bf (b)}
\resizebox{0.99\linewidth}{!}{%
\includegraphics{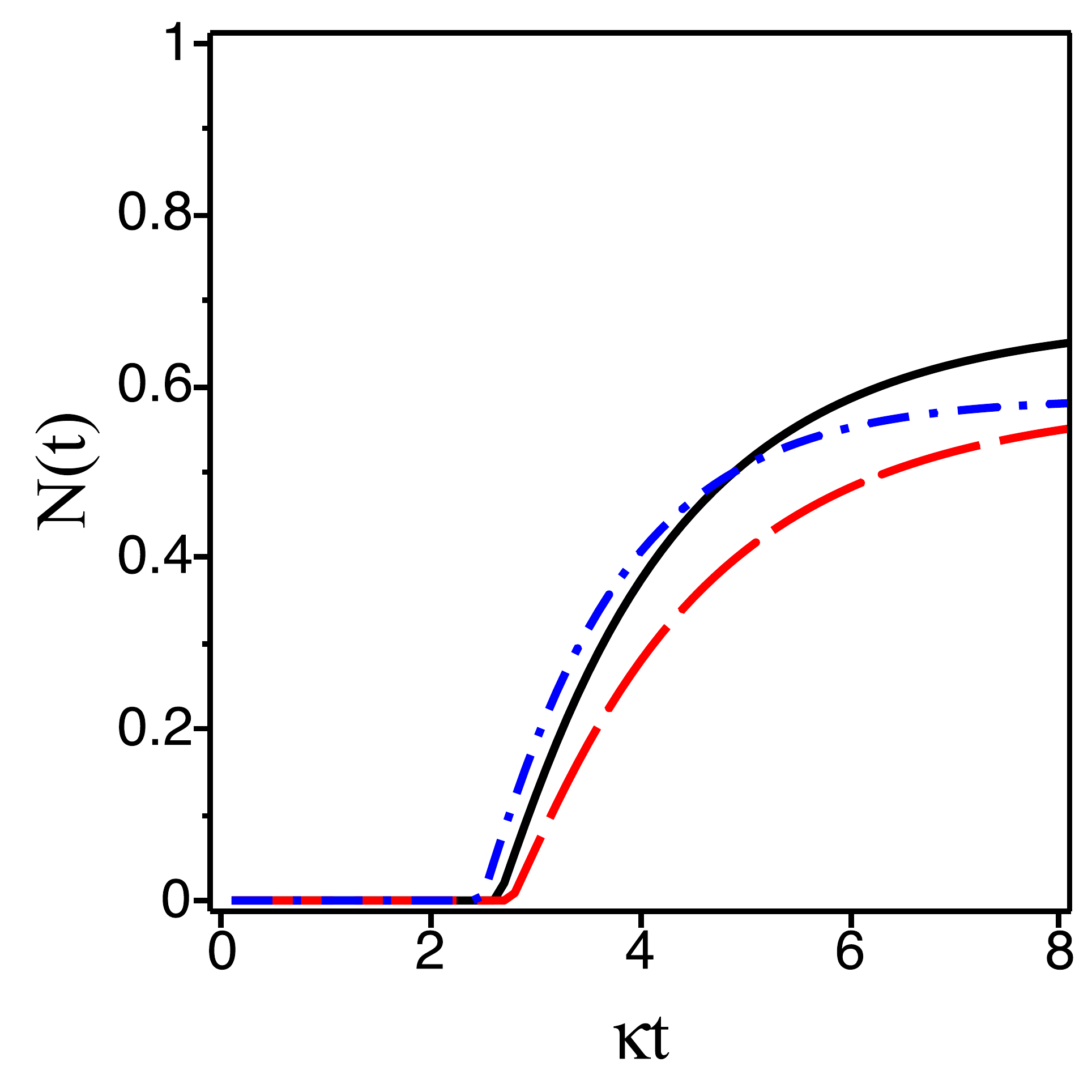}~\includegraphics{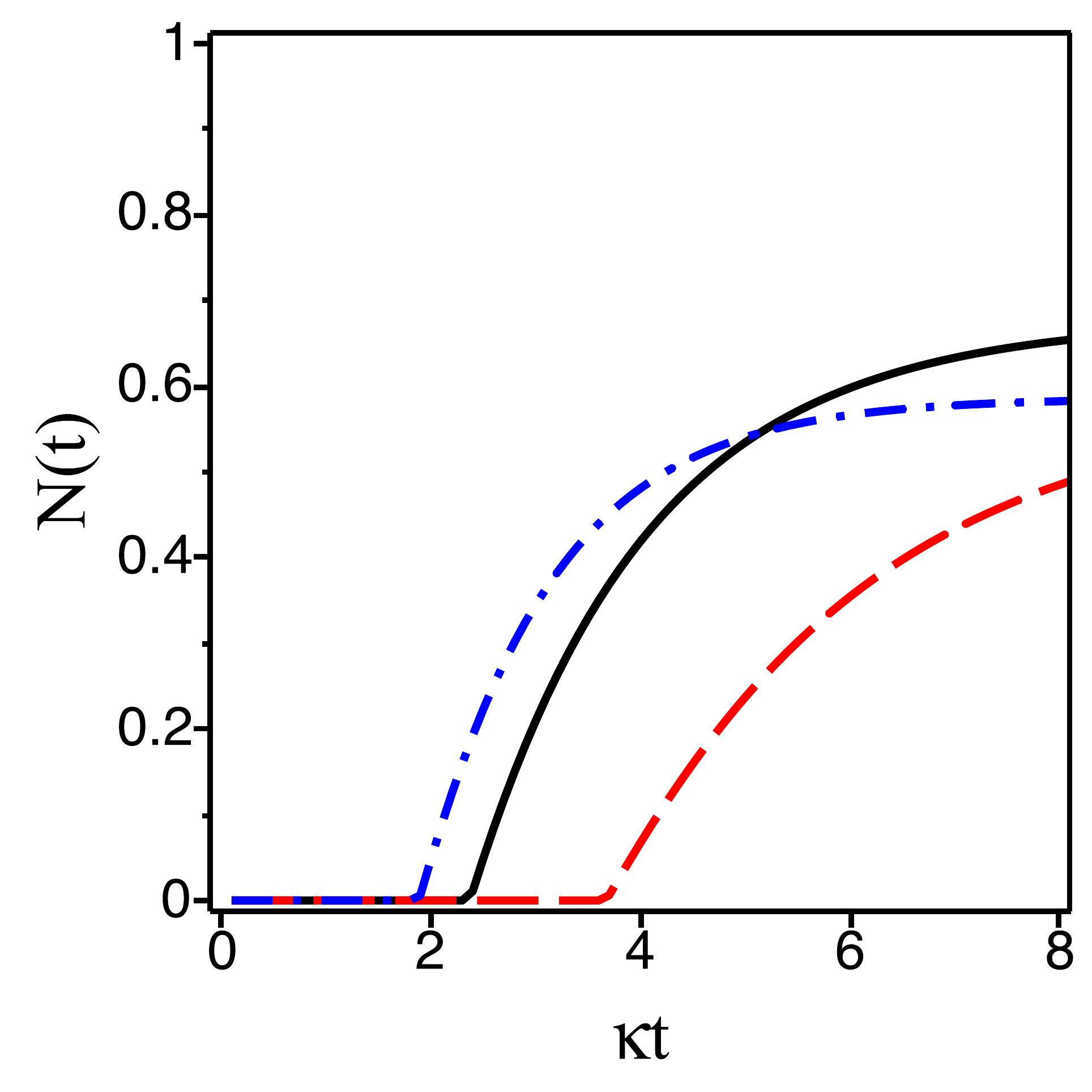}}
\caption{(Color online) Time build up of the logarithmic negativity for two different initial conditions and different damping rates with $N=0.125$, $|M|=\sqrt{N(N+1)}$, 
and $\kappa_{12}=\sqrt{\kappa_{1}\kappa_{2}}$. In frame (a) the initial condition is $\rho_{55}(0)=1$, and $\rho_{44}(0)=1$ in frame (b). 
Black solid line: $\kappa_1=\kappa_2=\kappa$, red dashed line: $\kappa_1=\kappa, \kappa_2=\frac{1}{2}\kappa $, and blue dash dotted line: 
$\kappa_1=\kappa, \kappa_2=2\kappa.$ }
\label{Fig5}
\end{figure}

One may notice from Fig.~\ref{Fig4} that the delay time of the transfer of the entanglement on cavities is not very sensitive to the number of photons initially present in the cavities. However, the delay time is sensitive to the damping rates of the cavities that the transfer could be further delayed when the cavities decay with different rates, $\kappa_{1}\neq \kappa_{2}$. This is illustrated in Fig.~\ref{Fig5} which shows the logarithmic negativity as a function of time for initial states $\ket{5}\equiv \ket{2_{A},0_{B}}$ and $\ket{4}\equiv\ket{1_{A},1_{B}}$, and for two different values of the ratio $\kappa_{2}/\kappa_{1}$. For the initial state $\ket 5$, the delay time is not increased further when $\kappa_{2}\neq\kappa_{1}$. On the other hand, for the initial state $\ket{4}$ and $\kappa_{1}\neq \kappa_{2}$, the delay time can be shortened or prolonged depending on whether $\kappa_{1}>\kappa_{2}$ or $\kappa_{1}<\kappa_{2}$. 

\begin{figure}[h]
\center{\includegraphics[width=0.6\linewidth]{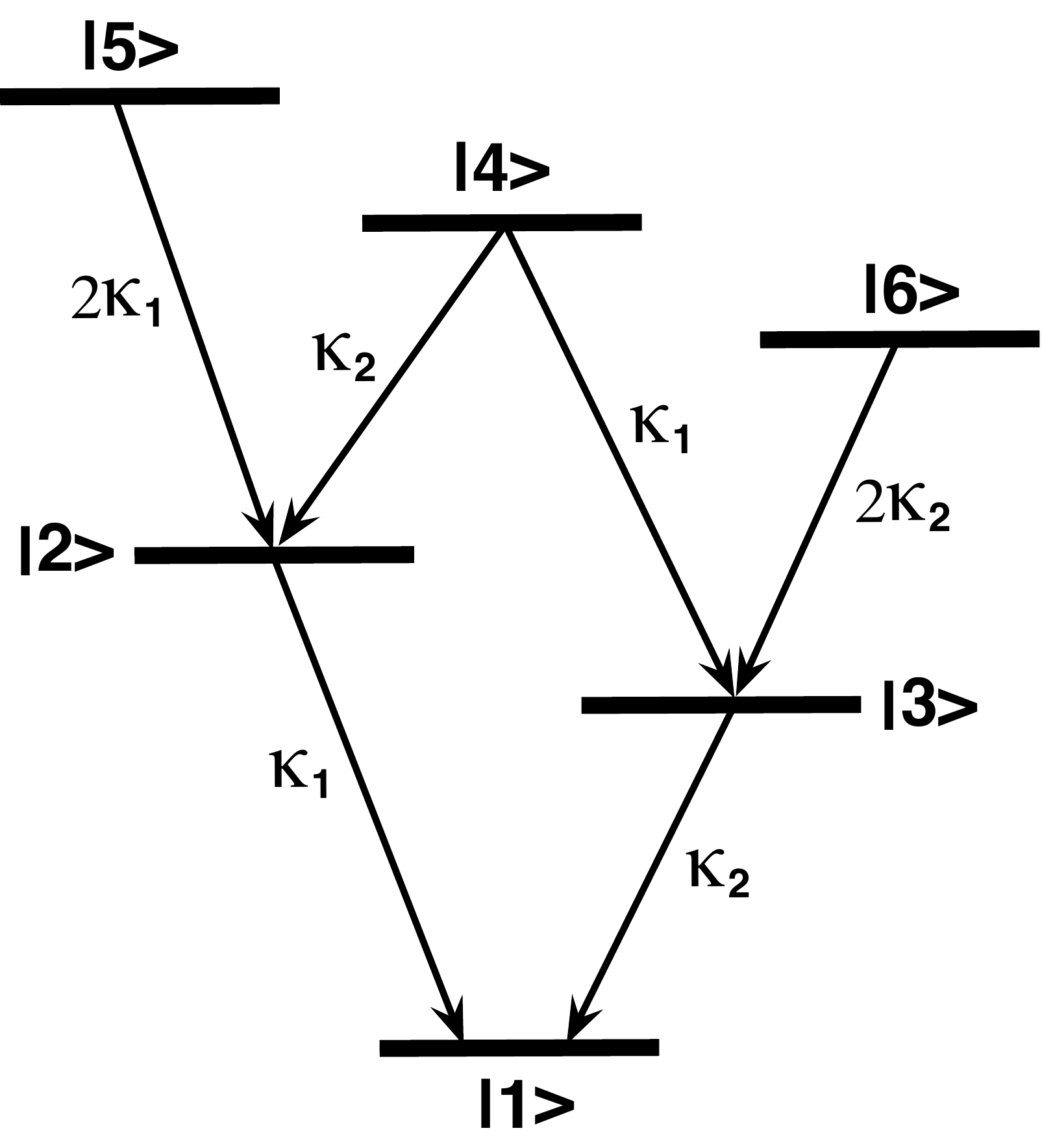}}
\caption{(Color online) Energy level diagram of the system of nondegenerate cavities with $\omega_{1}\gg\omega_{2}$ and transition rates between the energy states.}
\label{Fig6}
\end{figure}

The variation of the delayed time of the transfer of the entanglement with the ration $\kappa_{2}/\kappa_{1}$ can be readily understood if one refers to the energy-level diagram of the system, Fig.~\ref{Fig6}, which shows the allowed transitions between the energy levels and rates at which the excited states decay. There is a single pathway, with rate $2\kappa_{1}$ the state $\ket 5$ decays to the state $\ket 2$, but there are two pathways the state $\ket 4$ decays to the states $\ket 2$ and $\ket 3$, with the rates $\kappa_{2}$ and $\kappa_{1}$, respectively. 

Explanation of the variation of the delay time with the ratio $\kappa_{2}/\kappa_{1}$ for the initial state $\ket 4$ follows from the observation that increasing or decreasing of the damping rate $\kappa_{2}$ relative to $\kappa_{1}$ leads to an increase of the population of either $\ket 2$ or $\ket 3$ state, and consequently a longer decay time of the population of the single excitation states. 

\begin{figure}[h]
\hskip0.15cm {\bf (a)}\hskip4.0cm{\bf (b)}
\resizebox{0.99\linewidth}{!}{%
\includegraphics{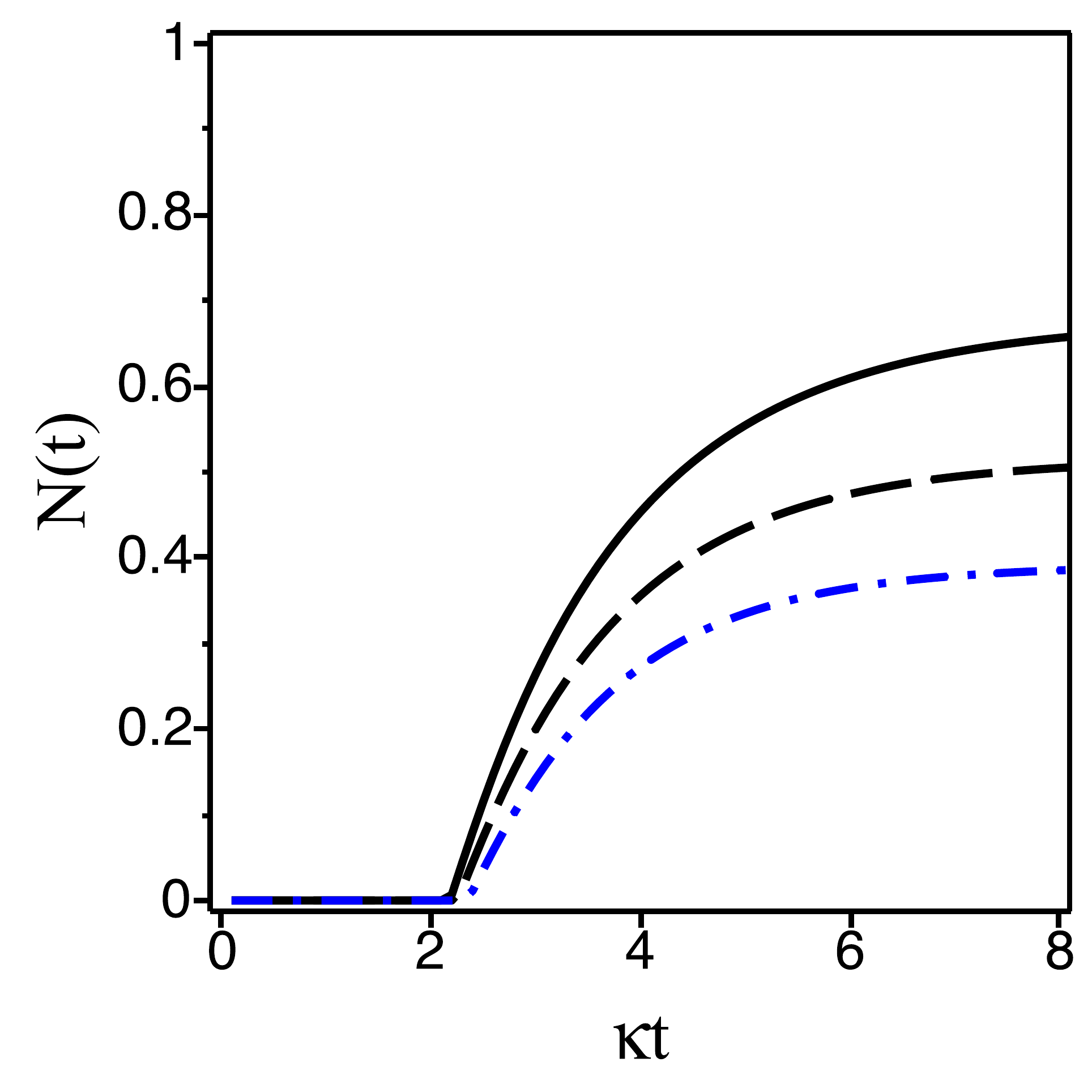}~\includegraphics{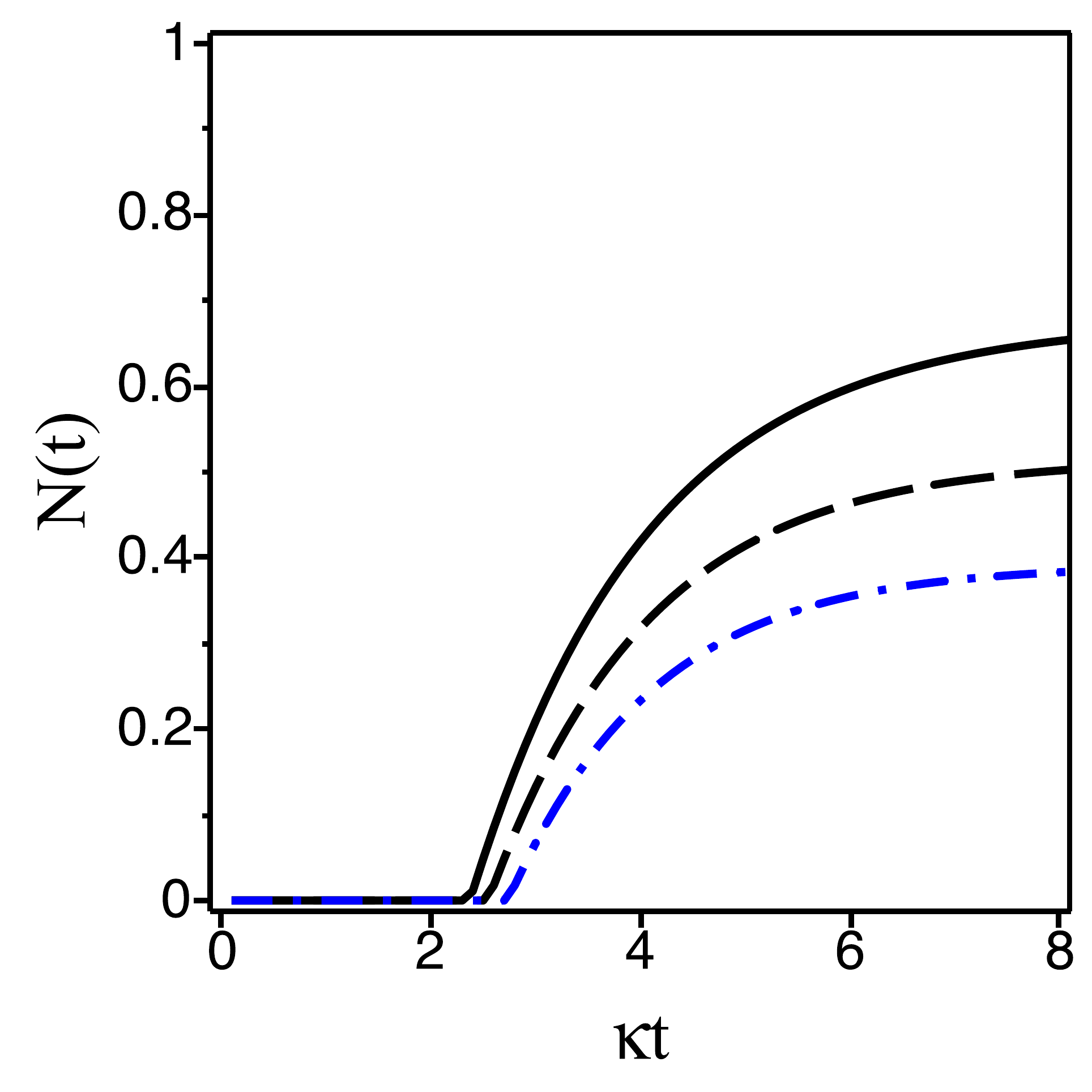}}
\caption{(Color online) Time build up of the logarithmic negativity for two different initial conditions and imperfect coupling $\eta<1$ with $N=0.125$, $|M|=\sqrt{N(N+1)}$, 
and $\kappa_1=\kappa_2=\kappa$. In frame (a) the initial condition is $\rho_{22}(0)=1$, and in frame (b) $\rho_{44}(0)=1$. In both frames $\eta =1$ (black solid line), $\eta=0.9$ (black dashed line), and $\eta=0.8$ (blue dashed-dotted line).}  
\label{Fig7}
\end{figure}

Finally, we address some practical limitations to the results presented in the above figures. In plotting the Figs.~\ref{Fig2}-\ref{Fig5}, we have assumed that the input squeezed field is perfectly coupled to the cavity modes, ie., $\eta=1$. However, in practice, the perfect coupling between the cavities and the input squeezed field could be difficult to achieve.
Therefore, in Fig.~\ref{Fig7}, we plot the logarithmic negativity for several values of the coupling efficiency $\eta$ and two different initial conditions $\rho_{22}(0)=1$, and $\rho_{44}(0)=1$. It is seen that the delay time is insensitive to the coupling efficiency. An imperfect coupling only affects  the amount of the transferred entanglement.

\section{Origin of the delayed mapping of entanglement}\label{sec4}

Here we present a qualitative understanding of the physical origin of the delayed mapping of entanglement. We will show that the delayed transfer of entanglement can be attributed to the presence of quantum jumps~\cite{vn95,mz96,gz04,hc08}. To do this, we rewrite the master equation (\ref{e8}) in terms of a coherent evolution governed by a non-Hermitian Hamiltonian and an incoherent evolution which is solely due to spontaneous emission events as
\begin{align}
\frac{d\rho}{dt} &=\frac{1}{i\hbar}\left[H_{\rm eff},\rho\right] + {\cal L}_{\rm sp} \rho ,\label{b21}
\end{align}
where
\begin{align}
H_{\rm eff} =  &  \ \hbar\!\sum_{j=A,B} \omega_{j}a_{j}^{\dagger}a_{j} \nonumber \\
&-\!\frac{1}{2}i\hbar \sum_{j=A,B} \kappa_{j}\left\{\left[N(\omega_{j})+1\right]a_{j}^{\dagger}a_{j} + N(\omega_{j})a_{j}a_{j}^{\dagger}\right\} \nonumber\\
& +\frac{1}{2}i\hbar \sum_{i\ne j=A,B}\kappa_{ij} \left[M(\omega_{j}) a_i^{\dagger} a_j^{\dagger} + M^{\ast}(\omega_{j}) a_i a_j\right] ,\label{b22}
\end{align}
represents a coherent nonunitary evolution of the system, and
\begin{align}
{\cal L}_{\rm sp}\rho = & \sum_{j=A,B}\kappa_{j}\left\{\left[N(\omega_{j})+1 \right]a_{j}\rho a_{j}^{\dagger}  + N(\omega_{j}) a_{j}^{\dagger}\rho a_{j} \right\} \nonumber \\
& -\sum_{i\neq j=A,B}\kappa_{ij} \left(M a_{i}^{\dagger}\rho a_{j}^{\dagger}  + M^{\ast}a_{i}\rho a_{j}\right) .\label{b23}
\end{align}
represents incoherent processes due to quantum jumps which contribute to the dynamics of the system resulting from a continuous measurement performed by the environment on the system.

Quantum jumps cause instantaneous switching between energy levels of the system, which changes the distribution of the population of the levels and coherences between them. For example, the quantum jumps change the two-photon coherence $\rho_{14}$ at a rate
\begin{align}
\left({\cal L}_{\rm sp}\rho\right)_{14} & = -\kappa_{12}M^{\ast} \left(\rho_{22}+\rho_{33}\right) ,\label{b28}
\end{align}
and the population of the state $\ket 4$ at a rate
\begin{align}
\left({\cal L}_{\rm sp}\rho\right)_{44} & = N\!\left(\kappa_{2}\rho_{22}+\kappa_{1}\rho_{33}\right) .\label{b29}
\end{align}
Clearly, in the presence of quantum jumps the two-photon coherence and the population of the state $\ket 4$ are leaking out through the one-photon states $\ket 2$ and $\ket 3$. Therefore, a significant reduction of the transfer of entanglement or no entanglement transfer are expected to be found at times the states $\ket 2$ and $\ket 3$ are significantly populated. 
From Eqs~(\ref{b22}) and~(\ref{b23}), it is worth noting that the populations in the $\ket 2$ and $\ket 3$ states is a consequence of single -photon incoherent processes.

Now we can ask ourselves what would happen with the transfer of entanglement if we ignore quantum jumps. 
Without quantum jumps the equations of motion (\ref{b5}) for the density matrix elements become
\begin{align}
\dot{\rho}_{11} =& -2N\kappa\rho_{11} + 2M\kappa\rho_{m} ,\nonumber\\
\dot{\rho}_{44} =& -2(N+1)\kappa\rho_{44} + 2M\kappa\rho_{m} ,\nonumber\\
\dot{\rho}_{ss} =& -\left(4N+1\right)\kappa\rho_{ss} +(N+1)\kappa\left(\rho_{44} +2\rho_{uu}\right) \nonumber\\
&+ N\kappa\rho_{11} -2M\kappa\rho_{m} ,\nonumber\\
\dot{\rho}_{uu} =& -2(N+1)\kappa\rho_{uu} +2N\kappa\rho_{ss} ,\nonumber\\
\dot{\rho}_{m} =& -(2N+1)\kappa\rho_{m} + M\kappa(\rho_{11}+\rho_{44}) ,\label{b26}
\end{align}
where 
\begin{align}
\rho_{ss} &= \frac{1}{2}\left(\rho_{22} +\rho_{33}\right) ,\quad \rho_{uu} =\frac{1}{2}\left(\rho_{55} +\rho_{66}\right) ,\label{b25}
\end{align}
are incoherent mixtures of the one-photon states $\ket 2$ and $\ket 3$ and the two-photon states $\ket 5$ and $\ket 6$, respectively.

It is seen that the evolution of the populations of the ground $\ket 1$ and the upper $\ket 4$ states is decoupled from the incoherent mixtures of the states $\ket 2$ and $\ket 3$. They are coupled to each other by a coherent two-photon excitation channel of the squeezed field. The strength of this coupling is proportional to the magnitude of two-photon
correlation ($M$) of the squeezed field. Clearly, the effect of quantum jumps is to introduce coupling between the superposition (entangled) and incoherent mixture states. 

In terms of the superposition states (\ref{b8}), the equations (\ref{b26}) take the form 
\begin{align}
\dot{\rho}_{\alpha\alpha} =& 0 ,\nonumber\\
\dot{\rho}_{\beta\beta} =& -2(2N+1)\kappa\rho_{\beta\beta} ,\nonumber\\
\dot{\rho}_{ss} =& -\left(4N+1\right)\kappa\rho_{ss} +2(N+1)\kappa\rho_{uu} \nonumber\\
&+2(2N+1)\kappa\rho_{\beta\beta} ,\nonumber\\
\dot{\rho}_{uu} =& -2(N+1)\kappa\rho_{uu} +2N\kappa\rho_{ss} .\label{b37}
\end{align}
The result $\dot{\rho}_{\alpha\alpha}=0$ indicates that the state does not evolves in time, ie. the state $\ket\alpha$ is a dark state. This suggests that this state if not initially 
populated it would never be populated. Thus, in the absence of the quantum jumps, the quantum correlations (entanglement) would never be transferred to the cavities.

\section{Conclusions}\label{sec5}

The transient response of single-mode cavities to the input squeezed vacuum field has been described. The response to the quantum correlations (entanglement) present in the squeezed field has been found to be a sensitive function of the initial conditions of the cavities. We have found that depending on the initial excitation of the cavities, the transfer 
of the quantum correlations can be delayed even though the absorption of photons from the field is not sensitive to the initial population.
In the case of empty cavities, with no initial excitation present, the transfer of the quantum correlations begins immediately after the squeezed field is turned on. In contrast, 
if the cavities are initially prepared in some of the excited states, the transfer is delayed by a finite time interval. The delayed time interval depends on the damping rates of the cavities and can be varied by varying the ratio of the damping rates. A detailed analysis has shown that the process of the delayed transfer of the quantum correlations is related to the presence of the population in the one-photon states of the cavity system. The transfer of the quantum correlations is postponed till the one-photon states of the system are almost completely depopulated. In other words, the system ”waits” for the population of the single-photon states to decay out before the quantum correlations start to build up between the cavities.

We have pointed out that the delay of the entanglement transfer can be understood as resulting from the presence of quantum jumps. We have shown that quantum jumps cause instantaneous switching between entangled and incoherent mixture states of the system, which changes the distribution of the population of the levels and coherences between them. Although the quantum jumps delay transfer of entanglement from the input field to the cavities, they are in fact needed for the transfer to occur. 

The results of our work can potentially be used in quantum communication schemes to control the transfer, or transmission time of entanglement or quantum states through a noisy channel, as well as to schemes involving qubits subjected to decoherence due to the coupling to the environment~\cite{dz12,mk14,rr15,jw16,bc19,mb20}. In such schemes, it is crucial that the transfer occurs is a short time to avoid decoherence. 
Our results show that the presence of population in the transmission qubits can impose restrictions on the transfer time of entanglement. 

\appendix

\section{Equations of motion for the density matrix elements}\label{App}

In this Appendix, we present explicitly the complete set of the equations of motion for the density matrix elements in the basis spanned by the product states (\ref{b1}). 
The products of operators appearing in the master equation (\ref{e2}) can be written as
\begin{align}
\hat{a}^{\dagger}_{A}\hat{a}_{A} &= \ket 2\bra 2 +\ket 4\bra 4 +2\ket 5\bra 5 ,\nonumber\\
\hat{a}_{A}\hat{a}^{\dagger}_{A} &= \ket 1\bra 1 +\ket 3\bra 3 +2\ket 2\bra 2 ,\nonumber\\
\hat{a}^{\dagger}_{B}\hat{a}_{B} &= \ket 3\bra 3 +\ket 4\bra 4 +2\ket 6\bra 6 ,\nonumber\\
\hat{a}_{B}\hat{a}^{\dagger}_{B} &= \ket 1\bra 1 +\ket 2\bra 2 +2\ket 3\bra 3 ,\nonumber\\
\hat{a}^{\dagger}_{A}\hat{a}_{B} &= \ket 2\bra 3 +\sqrt{2}\ket 5\bra 4 +\sqrt{2}\ket 4\bra 6 ,\nonumber\\
\hat{a}_{A}\hat{a}_{B} &= \ket 1\bra 4 ,\nonumber\\
\hat{a}^{\dagger}_{A}\hat{a}^{\dagger}_{B} &= \ket 4\bra 1 .\label{a1}
\end{align}

Using the state basis (\ref{b1}) and the representation (\ref{a1}) in the master equation (\ref{e8}), we find the following equations of motion for the populations and coherences
\begin{align}
\dot{\rho}_{11} =& -N\left(\kappa_{1}+\kappa_{2}\right)\rho_{11} +(N+1)\kappa_{1}\rho_{22} \nonumber\\
&+ (N+1)\kappa_{2} \rho_{33} + M\kappa_{12}(\rho_{14}+\rho_{41}) ,\nonumber\\
\dot{\rho}_{22} =& -\left[\left(3N+1\right)\kappa_{1} +N\kappa_{2}\right]\rho_{22} +(N+1)\kappa_{2}\rho_{44} \nonumber\\
&+N\kappa_{1}\rho_{11} +2(N+1)\kappa_{1}\rho_{55} -M\kappa_{12}(\rho_{14}+\rho_{41}) ,\nonumber\\
\dot{\rho}_{33} =& -\left[\left(3N+1\right)\kappa_{2} +N\kappa_{1}\right]\rho_{33} +(N+1)\kappa_{1}\rho_{44} \nonumber\\
&+N\kappa_{2}\rho_{11} +2(N+1)\kappa_{2}\rho_{66} -M\kappa_{12}(\rho_{14}+\rho_{41}) ,\nonumber\\
\dot{\rho}_{44} =& -(N+1)\left(\kappa_{1}+\kappa_{2}\right)\rho_{44} +N\kappa_{2}\rho_{22} + N\kappa_{1} \rho_{33} \nonumber\\
& + M\kappa_{12}(\rho_{14}+\rho_{41}) ,\nonumber\\
\dot{\rho}_{55} =& -2(N+1)\kappa_{1}\rho_{55} +2N\kappa_{1}\rho_{22} ,\nonumber\\
\dot{\rho}_{66} =& -2(N+1)\kappa_{2}\rho_{66} +2N\kappa_{2}\rho_{33} ,\nonumber\\
\dot{\rho}_{14} =& -\frac{1}{2}(2N+1)(\kappa_{1}+\kappa_{2})\rho_{14} \nonumber\\
&+ M\kappa_{12}(\rho_{11}+\rho_{44}-\rho_{22}-\rho_{33}) .\label{c4}
\end{align}

In the case of equal damping rates of the cavities, $\kappa_{1}=\kappa_{2}=\kappa_{12}=\kappa$, Eq.~(\ref{c4}) simplifies to
\begin{align}
\dot{\rho}_{11} =& -2N\kappa\rho_{11} +(N+1)\kappa\left(\rho_{22} +\rho_{33}\right) + M\kappa(\rho_{14}+\rho_{41}) ,\nonumber\\
\dot{\rho}_{22} =& -\left(4N+1\right)\kappa\rho_{22} +(N+1)\kappa\left(\rho_{44} +2\rho_{55}\right) +N\kappa\rho_{11} \nonumber\\
&- M\kappa(\rho_{14}+\rho_{41}) ,\nonumber\\
\dot{\rho}_{33} =& -\left(4N+1\right)\kappa\rho_{33} +(N+1)\kappa\left(\rho_{44} +2\rho_{66}\right) +N\kappa\rho_{11} \nonumber\\
&- M\kappa(\rho_{14}+\rho_{41}) ,\nonumber\\
\dot{\rho}_{44} =& -2(N+1)\kappa\rho_{44} +N\kappa\left(\rho_{22} +\rho_{33}\right) + M\kappa(\rho_{14}+\rho_{41}) ,\nonumber\\
\dot{\rho}_{55} =& -2(N+1)\kappa\rho_{55} +2N\kappa\rho_{22} ,\nonumber\\
\dot{\rho}_{66} =& -2(N+1)\kappa\rho_{66} +2N\kappa\rho_{33} ,\nonumber\\
\dot{\rho}_{14} =& -(2N+1)\kappa\rho_{14} + M\kappa(\rho_{11}+\rho_{44}-\rho_{22}-\rho_{33}) .\label{b5}
\end{align}

\end{document}